\documentclass[conference]{IEEEtran}

\usepackage{cite}
\usepackage{amsmath,amssymb,amsfonts}
\usepackage{algorithmic}
\usepackage{graphicx}
\usepackage{textcomp}
\usepackage{xcolor}
\usepackage{blkarray, bigstrut}
\usepackage{bm}
\usepackage{nicematrix}
\usepackage{enumerate}

\def\BibTeX{{\rm B\kern-.05em{\sc i\kern-.025em b}\kern-.08em
    T\kern-.1667em\lower.7ex\hbox{E}\kern-.125emX}}

\begin{document}
\bstctlcite{MyBSTcontrol}

\title{Multi-Target Tracking with Dependent Likelihood Structures in Labeled Random Finite Set Filters}

\author{\IEEEauthorblockN{Lingji Chen}
\IEEEauthorblockA{\textit{Motional} \\ Boston, USA \\
chen-lingji@ieee.org}}

\maketitle

\begin{abstract}
In multi-target tracking, a data association hypothesis assigns measurements to tracks, and the hypothesis likelihood (of the joint target-measurement associations) is used to compare among all hypotheses for truncation under a finite compute budget. It is often assumed however that an individual target-measurement association likelihood is independent of others, i.e., it remains the same in whichever hypothesis it belongs to. In the case of Track Oriented Multiple Hypothesis Tracking (TO-MHT), this leads to a parsimonious representation of the hypothesis space, with a maximum likelihood solution obtained through solving an Integer Linear Programming problem. In Labeled Random Finite Set (Labeled RFS) filters, this leads to an efficient way of obtaining the top ranked hypotheses through solving a ranked assignment problem using Murty's algorithm. In this paper we present a {\em Propose and Verify} approach for certain {\em Dependent Likelihood Structures}, such that the true hypothesis likelihood is evaluated jointly for the constituent track-measurement associations to account for dependence among them, but at the same time that ranking is still obtained efficiently. This is achieved by {\em proposing} a candidate ranking under an assumption of independence, and then {\em evaluating} the true likelihood one by one, which is guaranteed, for certain Dependent Likelihood Structures, to not increase from its candidate value, until the desired number of top ranked hypotheses are obtained. Examples of such Dependent Likelihood Structures include the Collision Likelihood Structure and the Occlusion Likelihood Structure, both encountered frequently in applications.
\end{abstract}


\IEEEpeerreviewmaketitle

\section{Introduction}

\begin{figure}[!htb]
 \centering
 \includegraphics[width=0.7\columnwidth,trim=150 20 200 20,clip]{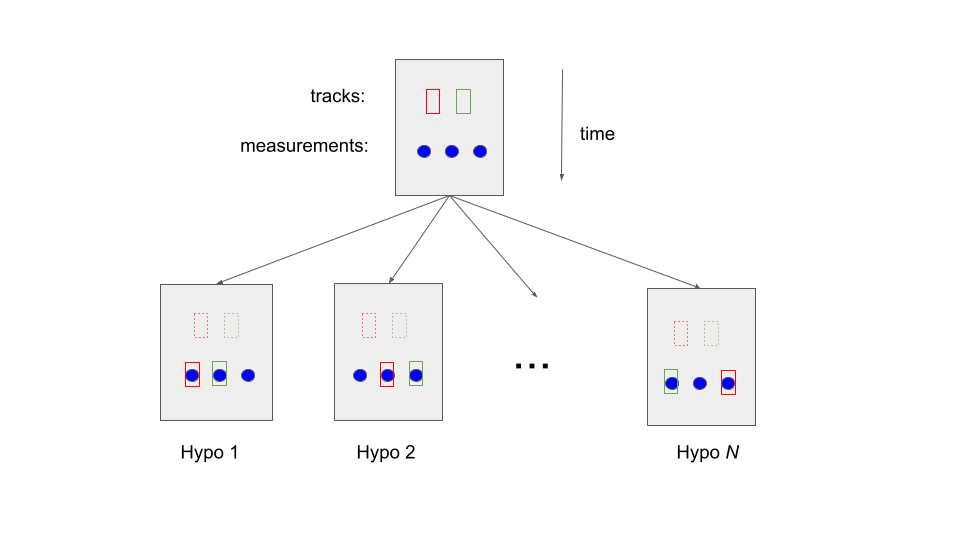} 
 \caption{An illustration of dependence. When considered in isolation, each red-red or green-green association is plausible. But when considered jointly, Hypo $N$ is almost impossible (for fast data rate).
}
\label{fig:intro}
\end{figure}

Multi-target 
tracking is a challenging problem that can be solved in the framework of Joint 
Probabilistic Data Association (JPDA), Multiple Hypothesis Tracking 
(MHT), and Random Finite Set (RFS); see a recent survey paper \cite{VoBN15_mtt} 
with a comprehensive list of references. In the latter two approaches, a data association hypothesis assigns measurements to tracks, and the hypothesis likelihood (of the joint target-measurement associations) is used to compare among all hypotheses for truncation under a finite compute budget. It is often assumed that an individual target-measurement association likelihood is independent of others, i.e., it remains the same in whichever hypothesis it belongs to. However, there are situations where the independence assumption does not hold, and any individual target-measurement association has to be evaluated jointly with all others in same hypothesis. Figure~\ref{fig:intro} illustrates this point, where the red and green  rectangles represent two targets, and the three blue dots represent the newly arrived frame of measurements. Many child hypotheses can be generated, e.g., both are on the left (Hypo 1), or both are on the right (Hypo 2). However, assuming that the sensor data rate is high, it is almost impossible for the two targets to swap positions in such a short time (Hypo $N$). But if examined in isolation, there is no order of magnitude difference in likelihoods among the individual, same-colored, dotted-solid associations in all hypotheses. This points to the need for evaluating likelihoods jointly in the context of the containing hypothesis, taking into account dependence among track-measurement associations. 

In the case of TO-MHT \cite{MorefieldC77}, assuming independence makes it possible to define a vector of individual track-measurement association costs, so that a global hypothesis likelihood is defined by a selection of entries in this vector satisfying certain constraints (that no two tracks share a measurement and no two measurements share a track). This leads to a parsimonious representation of the hypothesis space, and the maximum likelihood global hypothesis is obtained by solving an Integer Linear Programming problem. In such a framework, it is non-trivial to consider dependence as illustrated in Figure~\ref{fig:intro}.

In contrast, Labeled RFS filters \cite{VoBN14,VoB17}, as well as the historical Hypothesis Oriented Multiple Hypothesis Tracking (HO-MHT) \cite{ReidD79}, produce not only the best global hypothesis, but also a distribution over the global state in the form of a set of hypotheses with weights\footnote{that are informative and usually normalized but not necessarily true probabilities.}. For each of these explicitly expressed hypotheses, it may seem that its true likelihood can be computed with dependence taken into account, making the previous point moot. However, the challenge is in truncation. Due to the astronomical number of child hypotheses that are possible after receiving new measurements, truncation has to be carried out to keep only the top ranked hypotheses in computer memory. How to efficiently obtain such a ranking without exhaustively computing each hypothesis becomes the crux of the matter. Once again, assuming independence makes it possible to cast the ranking problem into a conventional ranked assignment problem that can be efficiently solved by Murty's algorithm \cite{MurtyK68} or its modern day incarnations \cite{LuQ18_murty}.

In this paper, we consider a special class of likelihood structures that allows us to adopt a {\em Propose and Verify} approach to ranking: We first assume independence and generate a list of candidate hypotheses in the decreasing order of likelihood, which can be carried out efficiently. Then we evaluate the true likelihood of each hypothesis in the list with the guarantee that the true value either stays the same as the candidate value used in the ranking proposal, or becomes lower. Thus we can both correctly and efficiently obtain the top $K$ hypotheses with their true likelihoods that take into account dependence. In practical terms, this means that to obtain the top $K=100$ hypotheses, we may be able to stop at say the 500th proposed candidate, without evaluating the true likelihoods of the rest of the hypotheses which can number in the billions. Two examples of the special structure are the Collision Likelihood Structure, where the true likelihood is the same if there is no collision, and zero otherwise; and the Occlusion Likelihood Structure, where the true likelihood is the same as when the probability of detection of a missed detection is given by an oracle, or lower.   

The paper is organized as follows.  Section~\ref{sec:pv} presents the main contribution of the Propose and Verify approach. Section~\ref{sec:app} describes how this approach is used in Generalized Labeled Multi Bernoulli (GLMB) filters. Section~\ref{sec:simu} illustrates the approach using a simple one dimensional example. A case for ubiquitous checking for collision/compatibility in extended target tracking is made in Section~\ref{sec:ext}, and conclusions are drawn in Section~\ref{sec:con}.

\section{Propose and Verify with special likelihood structures} \label{sec:pv}
In this section we present the main contribution of the paper.

\subsection{The convenience of independence}
The context is filtering with RFS. Let (labeled) RFS $X$ be the global target state, and $Z$ the global measurement. Then the filtering density $\pi$ is recursively obtained through set integrals \cite{VoBN14}:
\begin{align}
 \pi_k(X_k | Z_k) &= \frac{g_k(Z_k | X_k) \pi_{k|k-1}(X_{k-1})}{\int g_k(Z_k | X) \pi_{k|k-1}(X) \delta X}, \\
 \pi_{k+1|k}(X_{k+1}) &= \int f_{k+1|k}(X_{k+1}|X) \pi_k(X | Z_k) \delta X,
 \end{align}
where $f(\cdot)$ is the motion transition, and $g(\cdot)$ is the measurement likelihood.

Given a set of measurements $Z=\{z_j, j=1, \ldots, |Z|\}$, a data association hypothesis $\theta$ for a labeled RFS $X$ assigns to each label $\ell$ either a measurement index in a one-to-one fashion, or the special index 0 to denote a missed detection. Conditional independence is often assumed that leads to the following factored form of the likelihood function \cite{VoB17}
\begin{equation}
 g(Z|X) \propto \sum_{\theta: {\rm valid}} \prod_{(x, \ell) \in X} \psi_Z^{(\theta(\ell))}(x, \ell),
\end{equation}
where 
\begin{equation} \label{eqn:psi}
\psi_Z^{(j)}(x, \ell) = \left\{ \begin{array}{cl} \frac{P_D(x, \ell) g(z_j|x, \ell)}{\kappa(z_j)}, & \mbox{ if } j \in \{1, \ldots, |Z|\} \\ 1 - P_D(x, \ell), & \mbox{ if } j = 0 \end{array} \right.  ,
\end{equation}
in which $\kappa(\cdot)$ denotes clutter density, and $P_D(\cdot)$ probability of detection.

It can be seen from the above two equations that, once the label $\ell$ is assigned the measurement $\theta(\ell)$, its likelihood of association is fixed; it ``does not care'' how other labels in the same hypothesis are assigned. The factored form has two advantages:
\begin{enumerate}
 \item The update of the kinematic distribution of each track is carried out only with ``its own'' measurement, and
 \item Taking the negative log, the ``cost'' of a hypothesis can be defined as a sum of selection from a cost matrix, whose entries are constructed independent of the hypotheses. 
\end{enumerate}

The second benefit leads to a ranking problem that can be efficiently solved using Murty's algorithm. However, there are likelihood structures that can consider the interactions among the tracks but still keep a factored form, hence reaping the first benefit listed above. What we need then, is an efficient ranking algorithm for such structures. 

\subsection{The factored form that accounts for dependence}
We introduce for each track $(x, \ell)$ a factor  $\lambda(X, \theta, \ell)$ such that the likelihood function takes the form 
\begin{equation}
 g(Z|X) \propto \sum_{\theta: {\rm valid}} \prod_{(x, \ell) \in X}  \lambda(X, \theta, \ell) \psi_Z^{(\theta(\ell))}(x, \ell).
\end{equation}
Then within a given hypothesis, each track can be made aware of ``what's going on'' with respect to other tracks, and adjust its own likelihood term accordingly. This can account for not all but certain dependence among the tracks, while still maintaining a factored form so that the posterior distribution of each track can be computed separately.

The following are two examples of such likelihood structures.

\subsubsection{The Collision Likelihood Structure}
The term $\psi_Z^{(\theta(\ell))}(x, \ell)$ is defined as in Equation~(\ref{eqn:psi}). Let 
\[ 
{\tt collide}(X, \theta)
\]
denote a checking procedure that returns {\tt true} if the hypothesis under consideration induces a collision among the tracks, as is illustrated by Hypo $N$ in Figure~\ref{fig:intro}. Then the additional factor is defined as 
\begin{equation}
 \lambda(X, \theta, \ell) = \left\{ \begin{array}{rl} 0, & \mbox{ if } \mbox{\tt \footnotesize collide}(X, \theta), \\
 1, & \mbox{ otherwise}. \end{array} \right. . 
\end{equation}
In other words, when the hypothesis $\theta$ is actually evaluated, and a collision is found, then it is dropped from further consideration. Otherwise, the likelihood is the same as in the case when independence is assumed.

\subsubsection{The Occlusion Likelihood Structure}
The term $\psi_Z^{(\theta(\ell))}(x, \ell)$ is modified as follows:
\begin{equation} \label{eqn:psi2}
\psi_Z^{(j)}(x, \ell) = \left\{ \begin{array}{cl} \frac{P_D(x, \ell) g(z_j|x, \ell)}{\kappa(z_j)}, & \mbox{ if } j \in \{1, \ldots, |Z|\} \\ 1, & \mbox{ if } j = 0 \end{array} \right.  .
\end{equation}
In other words, if a track has a missed detection, then it is optimistically assumed that the track is under some occlusion and therefore cannot be seen. Let 
\[
{\tt occluded}(X, \theta, \ell)  
\]
be a checking procedure that returns {\tt true} if track $\ell$ is occluded in this hypothesis. Then the additional factor is defined as 
\begin{equation}
 \lambda(X, \theta, \ell) = \left\{ \begin{array}{cl} 1 - P_D(x, \ell), & \mbox{ if } j=0 \mbox{ and} \\ &  \mbox{ not {\tt \footnotesize occluded}}(X, \theta, \ell), \\
 1, & \mbox{ otherwise}. \end{array} \right. . 
\end{equation}
This means that when a hypothesis is actually evaluated, the ``benefit of the doubt'' previously given to a missed detection can now be verified: If it is not warranted, then the likelihood has to be reduced to the true value. Otherwise, it is the same as in the case when independence is assumed and an oracle gives the correct probability of detection for missed detections.

\subsection{The Propose and Verify algorithm} \label{sec:pvalgo}
One crucial point to note is that 
\begin{equation}
 0 \le \lambda(X, \theta, \ell) \le 1.
\end{equation}
This gives us an efficient way to obtain ranked hypotheses, through a Propose and Verify process:
\begin{enumerate}
 \item Set $\lambda(X, \theta, \ell) = 1$ and construct, using Murty's algorithm, an {\tt iterator} over the hypotheses in descending order of optimistic likelihoods. This means listing the hypotheses through lazy evaluation, perhaps using method calls such as {\tt has\_next()} and {\tt get\_next()} \cite{ChenL18_code}.
 \item Evaluate each hypothesis and calculate the values of $\lambda(X, \theta, \ell)$. Insert the hypothesis into a separate sorted list {\tt result} in descending order of true likelihoods.
 \item Whenever 
 \begin{equation}
 \prod_{(x, \ell) \in X} \lambda(X, \theta, \ell) = 1
 \end{equation}
 happens, the true likelihood is the same as the optimistic likelihood, and therefore the ranking of this hypothesis and those before it in {\tt result} is verified (because new entries can never go before it).
 \item Stop when the desired number of top $K$ hypotheses are obtained, or when no more hypotheses are available.
\end{enumerate}

\begin{figure}[!htb]
 \centering
 \includegraphics[width=0.99\columnwidth,trim=30 80 50 20,clip]{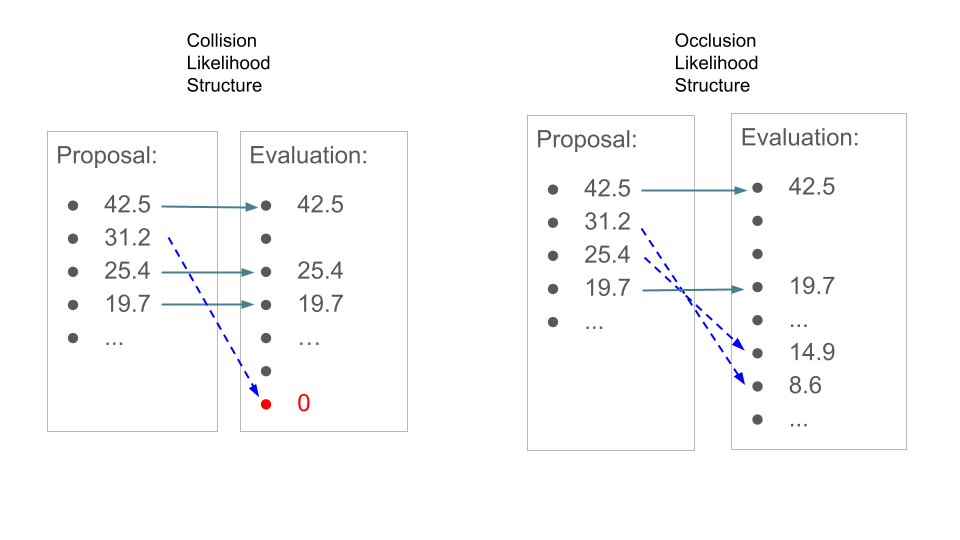} 
 \caption{Illustration of the Propose and Verify process for the Collision Likelihood Structure (left) and the Occlusion Likelihood Structure (right). The evaluated true likelihood is guaranteed to not increase; any ``stays the same'' hypo provides a certificate of ranking for itself and those above.}
\label{fig:structures}
\end{figure}

The algorithm is pictorially demonstrated in Figure~\ref{fig:structures}.

\section{Application in $\delta$-GLMB filters} \label{sec:app}
For completeness, we summarize how $\delta$-GLMB filtering is carried out with the likelihood structures described above. In the proposal stage, the main steps are the same as given in \cite{VoB17}, where measurement update is performed jointly with  
motion prediction. For each hypothesis, there 
is an associated Labeled Multi-Bernoulli (LMB) birth model; by treating birth probability as survival 
probability ``from nothing,'' we can treat both existing tracks and newborn 
track candidates in the same way, and refer to them simply as tracks. 

To determine the most likely ways of associating tracks with measurements, we 
construct a likelihood matrix that has rows for tracks, and three column blocks 
for measurements, missed detections, and deaths respectively. 
This matrix layout is shown here in Table~\ref{tab:layout} for easy reference; 
it is a simplified version of Figure 1 in \cite{VoB17}, before taking the logarithm of the  
likelihood ratios (over clutter density). 

\begin{table*}
\centering
\begin{blockarray}{cccccccccc}
 ~ & ~ & \mbox{detected} & ~ & ~ & \mbox{missed} & ~ & ~ & \mbox{died} & ~  \\
~ &$z_1$ &  $z_j$ &$z_M$ & $\bar{T}_1$ & $\bar{T}_i$ & $\bar{T}_N$ & 
$\hat{T}_1$ & $\hat{T}_i$ & $\hat{T}_N$\\
\begin{block}{c[ccc|ccc|ccc]}
\bigstrut[t]
$T_1$  & $\star$ & $\star$ & $\star$ & $\star$ & $0$ & $0$ & $\star$ 
& $0$ & $0$ \\
$T_i$  & $\star$ & $\eta_i(j)$ & $\star$ & $0$ & $\eta_i(0)$ & $0$ & 
$0$ & $\eta_i(-1)$ & $0$ \\
$T_N$  & $\star$ & $\star$ & $\star$ & $0$ & $0$ & $\star$ & $0$ 
& $0$ & $\star$ \bigstrut[b]\\
\end{block}
\end{blockarray}
\protect\caption{Likelihood ratio matrix layout (a simplified version of Figure 1 in \cite{VoB17})} \label{tab:layout}
\end{table*}

An entry in the first 
column block is the likelihood of a track having survived and being observed by 
that measurement, divided by the density of the measurement as clutter. 
The second column block is diagonal, and an entry on it is the probability of a track 
having survived but being mis-detected. The third column block is also diagonal, 
and an entry on it is the probability of a track having died. We take the negative 
log of the likelihood ratio matrix to get a cost matrix. A valid data association is 
defined by an assignment of the matrix such that each row has {\em exactly one} entry 
selected, and each column has {\em at most one} entry selected. The sum of the 
selected entries defines the cost of the hypothesis, the smaller the cost, the 
higher the likelihood.  

The best assignment can be found by using the Munkres algorithm 
\cite{MunkresJ57_algorithms}, while the best $K$ assignments  can be enumerated 
by using the Murty's algorithm \cite{MurtyK68}. Both have modern, faster 
versions; see \cite{LuQ18_murty} and the references therein. Since all current 
hypotheses perform this operation, and the union of their children constitute 
the next generation of hypotheses, a suboptimal but parallelizable selection 
scheme is to allocate, a priori, fixed number of children for each hypothesis, 
e.g., in proportion to its prior weight. The scheme is suboptimal because it 
may turn out that some child of a high-weight parent has a smaller weight than 
some would-be child of a lower-weight parent if the latter were given a larger 
allocation. 

If we implement the Murty's algorithm in the style of an {\tt 
iterator}, i.e., with methods such as {\tt has\_next()} and {\tt get\_next()}, 
then the optimal selection scheme\footnote{This idea was first proposed to the 
author by Peter Kingston.}  can be defined as follows in a round robin fashion:
\begin{enumerate}
 \item Let each hypothesis calculate and {\tt pop} its best child; this is the top entry in the {\tt result} list described in Section~\ref{sec:pvalgo}. Put these popped children in a selection 
{\tt buffer}.
 \item Copy the best out of the {\tt buffer}, and replace the content in this spot with the next 
best child from the same parent.
 \item Repeat until all top $K$ hypotheses have been obtained, or until no more 
children are available.
\end{enumerate}

It is worth noting that each parent hypothesis has the freedom to choose how to calculate its best child. When collision is of concern, it can adopt the Collision Likelihood Structure and use the Propose and Verify procedure. When occlusion is of concern, the Occlusion Likelihood Structure can be used. When independence can reasonably be assumed, then the plain vanilla algorithm can be used for its lower computational cost.

\section{Some illustrations} \label{sec:simu}
We use a simple, one-dimensional simulation to illustrate the difference our proposed approach can make. This is motivated by a statement made in the Conclusions section in \cite{ChongC18_mht}:
\begin{quotation}
 Most MHT algorithms assume targets have i.i.d. motion models. The independence assumption is violated
when targets move as a group, or when vehicles are moving on a single-lane road. Exploiting this dependence should improve association performance.
\end{quotation}

\begin{figure}[!htb]
 \centering
 \includegraphics[width=0.99\columnwidth,trim=80 5 100 5,clip]{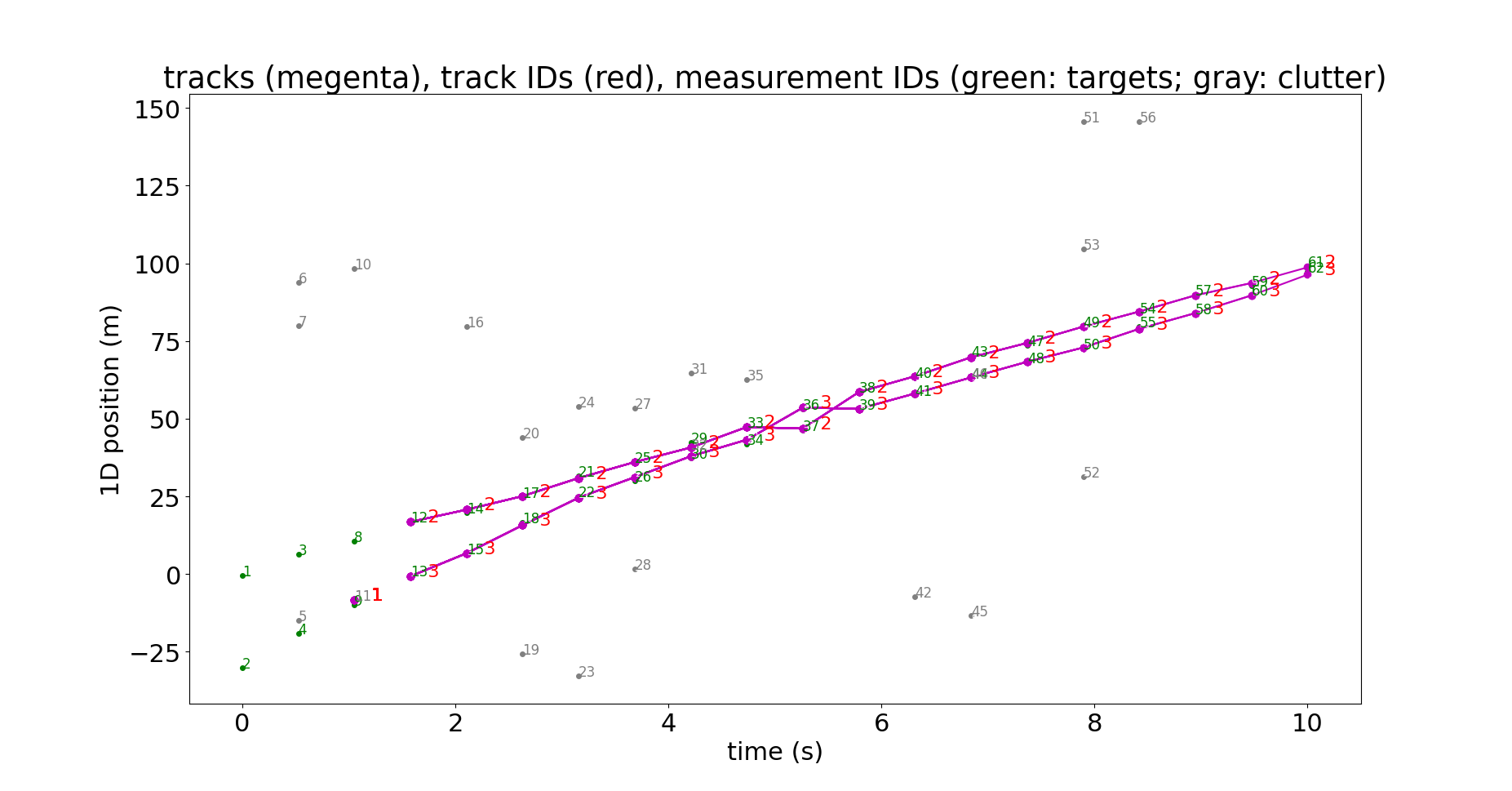} %
 \caption{A one-dimensional simulation of targets on a single lane road. Assuming independence of data assignment, some Monte Carlo runs produce track histories that indicate that one target has overtaken the other, which should not happen.}
\label{fig:simu1}
\end{figure}

Two targets, modeled as two points, travel on a one-lane road modeled as a line. However, we assume that the targets have an extent, and therefore one cannot get arbitrarily close to or pass the other. A position sensor produces the measurements including clutter, and a $\delta$-GLMB filter does the tracking. The simulation and filtering set up is given in Appendix~\ref{appd:setup}; it is along similar lines as in \cite{ChenL21_merge}, which provides detailed description of the ``merge and split'' version of $\delta$-GLMB used in this simulation. At each iteration, the best state estimates are defined, for simplicity, as the state estimates from the highest weighted hypothesis. Assuming independence, and without specifically addressing the issue that one target cannot overtake the other in this one-lane scenario, the tracks in some Monte Carlo simulations can look like those shown in Figure~\ref{fig:simu1}.

\begin{figure}[!htb]
 \centering
 \includegraphics[width=0.99\columnwidth,trim=80 5 100 5,clip]{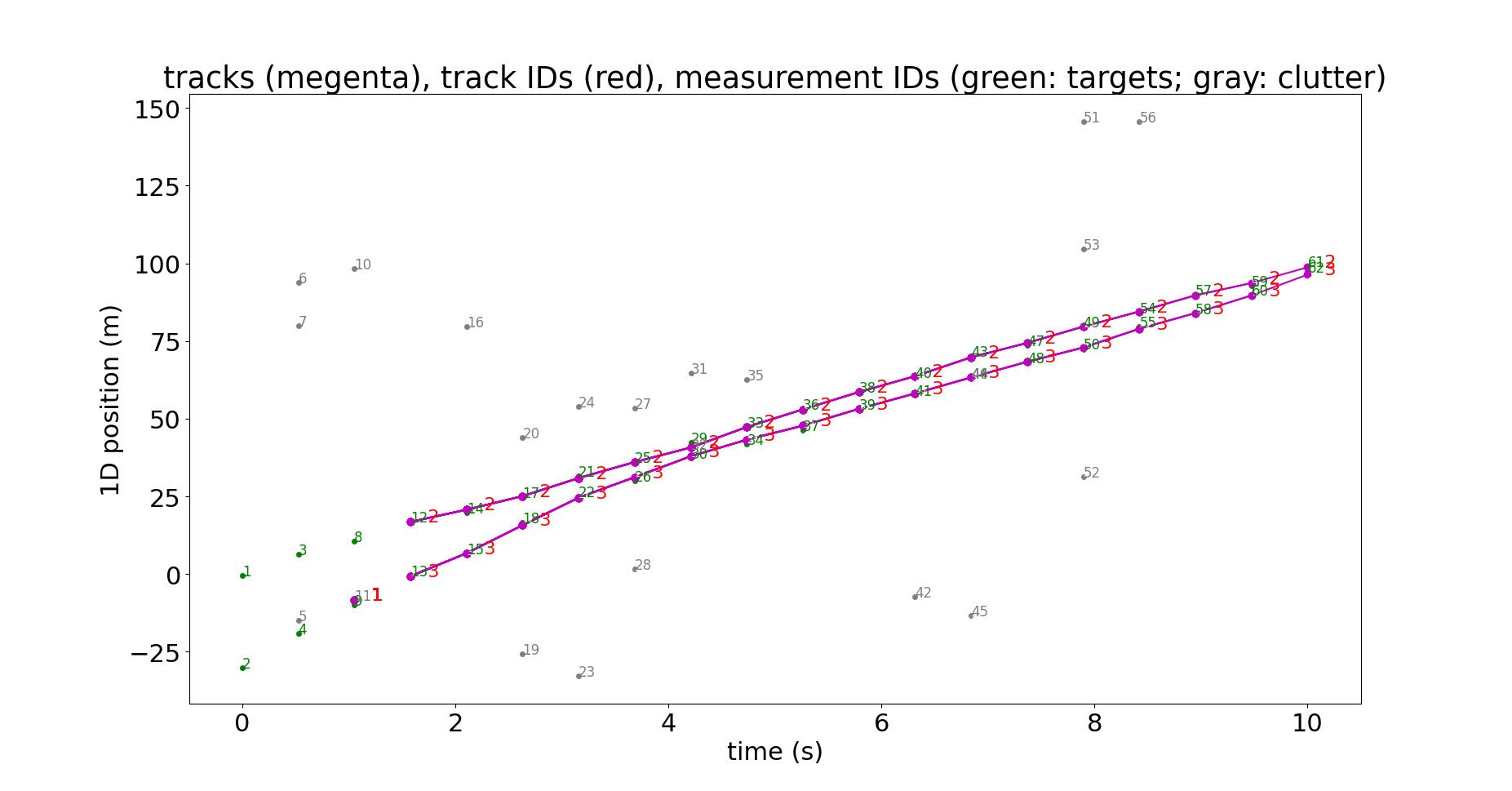} 
 \caption{A one-dimensional simulation of targets on a single lane road. Adopting the Collision Likelihood Structure and the Propose and Verify algorithm, targets follow each other without taking over each other.}
\label{fig:simu2}
\end{figure}

However, if we adopt the Collision Likelihood Structure and the Propose and Verify algorithm, then the track trajectories are feasible, as shown in Figure~\ref{fig:simu2}.

In this one dimensional scenario that is used for illustration, we assume that the extent of a target is not negligible but can be quite small. This enables us to define collision only as a result of overtaking, which is determined by the following criteria:
\begin{enumerate}[(i)]
 \item Target IDs that are in the parent hypothesis but not in the child hypothesis (target deaths) do not cause collision.
 \item Targets IDs that are in the child hypothesis but not in the parent hypothesis (target births) do not cause collision.
 \item Target IDs that are in both hypotheses (persistent targets), when sorted by road position, have to maintain the same order from parent hypo to child hypo; otherwise a collision has occurred. 
\end{enumerate}

There is some subtlety though. The track histories we have shown are not true track trajectories in the sense of \cite{GarciaFernandezA20_multiple}. We have simply connected the best state estimates across time using their IDs (labels), which may be different from the pedigrees of the current best hypo \cite{ChenL18}. For the latter, a collision-free transition from a parent hypo to a child hypo guarantees that the trajectory defined by the pedigree is collision-free. But for the former, if there is a ``hypo-hop\cite{ChenL18},'' then there is no such guarantee.  

\begin{figure*}
 \centering
 
 \includegraphics[height=0.99\textheight]{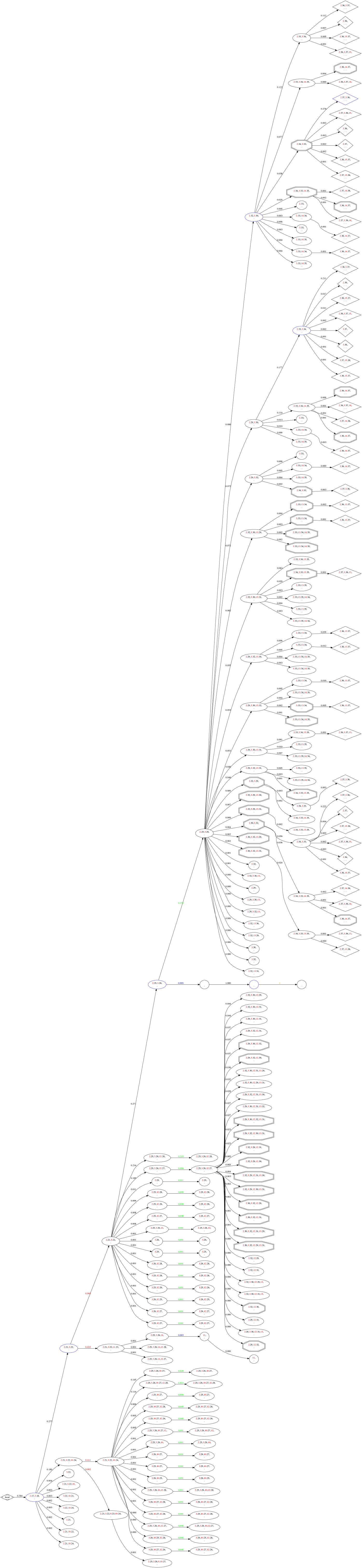} \ \ 
 \fbox{
 \includegraphics[height=0.99\textheight,trim=1700 9000 0 120,clip]{figures/twocars11.png}
}
 \caption{Evolution of hypotheses (left) and a zoomed-in detail (right) when running in debug mode and assuming {\em independence}. Each ellipse/diamond/double-octagon represents a hypo, with red track IDs and black measurement IDs within, and weights on the arc. Blue border indicates the best hypo (one for each factor \cite{ChenL21_merge}) at that time. A double-octagon hypo is one that {\em would} be declared a collision according to the criteria listed above, even though it is not forbidden. There are some hypo-hops in this example. }
\label{fig:tree}
\end{figure*}

To see the above statement in some details, let us run the simulation where we assume independence in debug mode, in order to show the evolution of hypotheses. Even though we do not forbid taking over in data association, for the purpose of comparison we apply the criteria described earlier to identify situations where taking over does occur. Figure~\ref{fig:tree} includes a full ``tree'' plot of the evolution of hypotheses on the left. The full explanation of the notations, including the Merge/Split algorithm, is given in \cite{ChenL21_merge} and presented in Appendix~\ref{appd:notation} for ease of reference. A zoomed-in panel is presented on the right in Figure~\ref{fig:tree} to show where hypo-hop and taking over occur.

\begin{figure}[!htb]
 \centering
 \includegraphics[width=0.99\columnwidth,trim=80 5 100 5,clip]{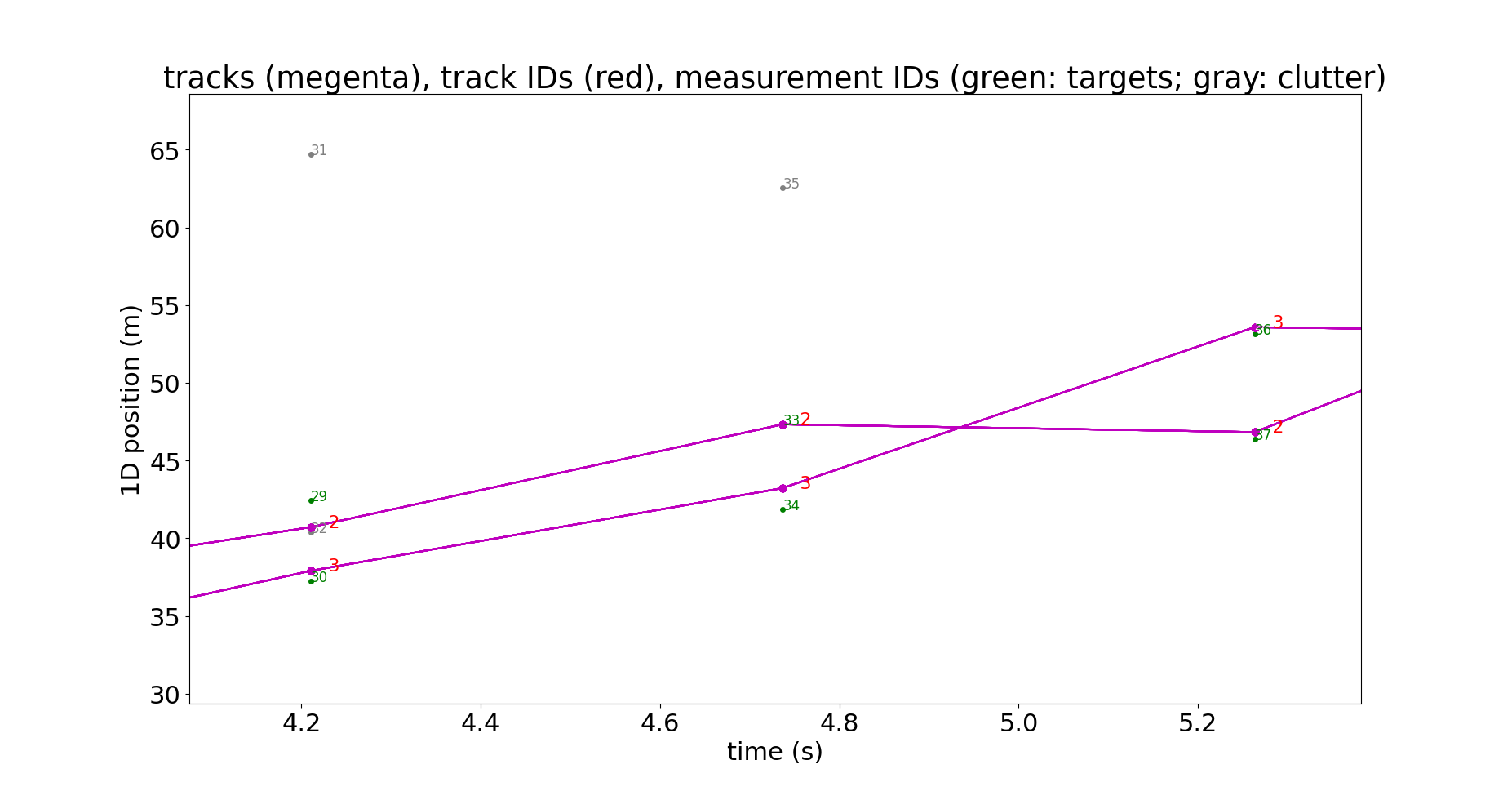} 
 \caption{A zoomed-in portion of Figure~\ref{fig:simu1}.}
\label{fig:zoomed}
\end{figure}

What is relevant for our current discussions is as follows:  Each ellipse/diamond/double-octagon represents a hypo, within which are listings of red track IDs followed by black measurement IDs, and with weights on the incoming arc. Blue border indicates the best hypo (one for each factor \cite{ChenL21_merge}) at that frame time. For debugging purposes, we apply the above collision criteria to each parent-child hypo pair, and set a flag value for each child that would cause a collision, even though we are not doing anything about it in this ``assuming independence'' run. Such a child is shown as a double-octagon hypo in the diagram. Note that there are some hypo-hops in this example: The best, blue hypo on the upper right corner did not come from a blue parent. For ease of cross referencing, a zoomed-in portion of Figure~\ref{fig:simu2} is shown in Figure~\ref{fig:zoomed}. It would appear that the child hypo that assigns Measurements 36 and 37 is a ``collision'' one, but it is actually not due to hypo-hop.

\section{Related work} \label{sec:related}
Separable multi-object likelihood functions are introduced in \cite{VoB10_joint}, and are used in \cite{PapiF15_generalized} to construct a tracktable, approximate distribution to handle generic multi-object likelihood functions. It is unclear how these proposed frameworks help solve the problem illustrated in Figure~\ref{fig:intro}. Another paper \cite{LiS18_multiobject} that handles generic observation models only gives examples that assume some form of conditional independence that would not hold for the situation in Figure~\ref{fig:intro}. 

\section{A call to arms} \label{sec:ext} 
Going from point target tracking to extended target tracking is a giant step. Not only do we have to account for the estimation of target extent which is often coupled with target dynamics, but also we have to contend with the challenge that it is no longer the case that one target gives rise to at most one measurement in a given data frame. See \cite{GranstromK17_jaif} for a recent overview. 

\begin{figure}[!htb]
 \centering
 \includegraphics[width=0.75\columnwidth]{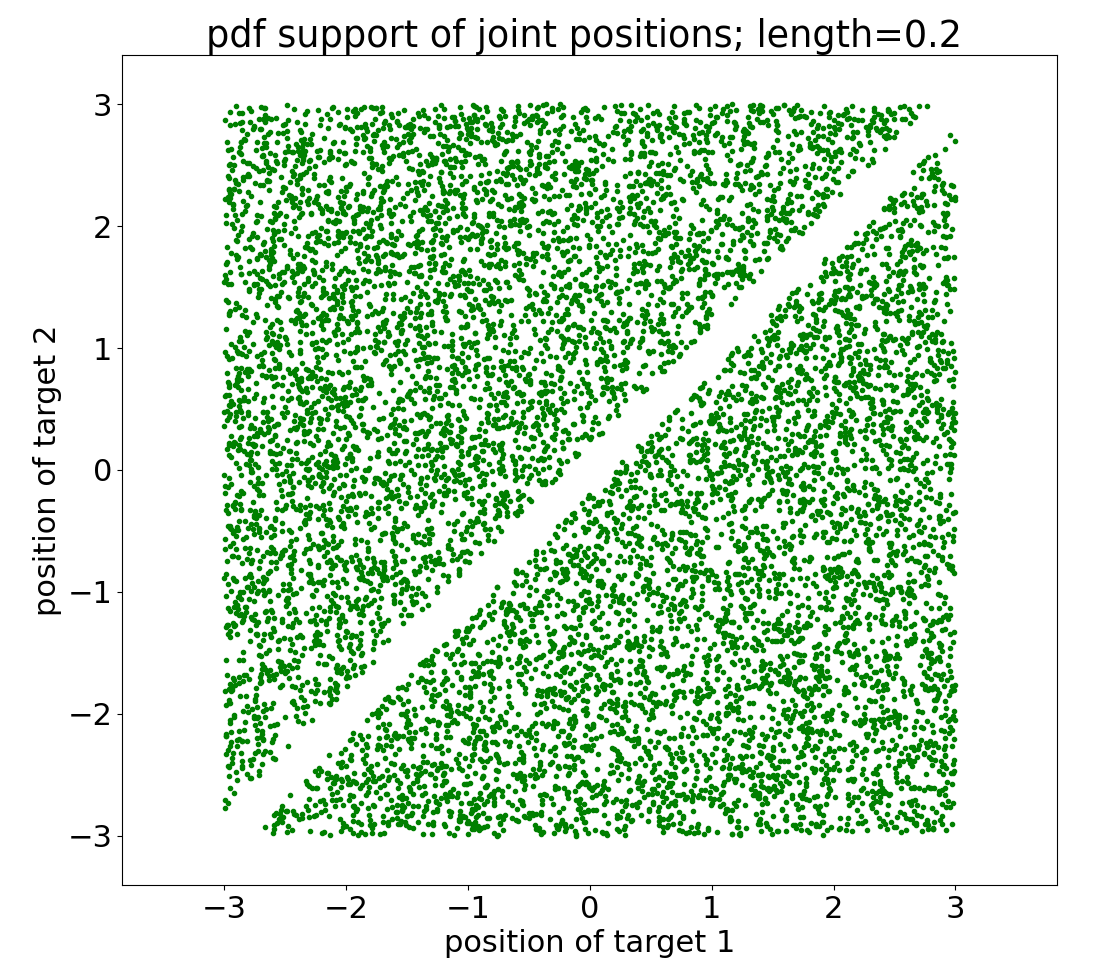} 
 \caption{The support of the pdf of the joint positions of two targets with a fixed length. Axes are limited between -3 and 3.}
\label{fig:twopos}
\end{figure}

However, there is another, more fundamental challenge that has not, in our view, received adequate attention in the tracking community. The challenge is the lack of new mathematical tools to deal with representation and inference when {\em spatial compatibility} is of essential concern. To illustrate this point, consider a target whose state consists only of a one-dimensional position (of its center), and a length. If we fix the length of all targets at a given constant, and consider the joint distribution of the positions of two targets, then the support of this probability density function (pdf) will look like that shown in Figure~\ref{fig:twopos}.  

\begin{figure}[!htb]
 \centering
 \includegraphics[width=0.99\columnwidth,trim=10 0 10 0,clip]{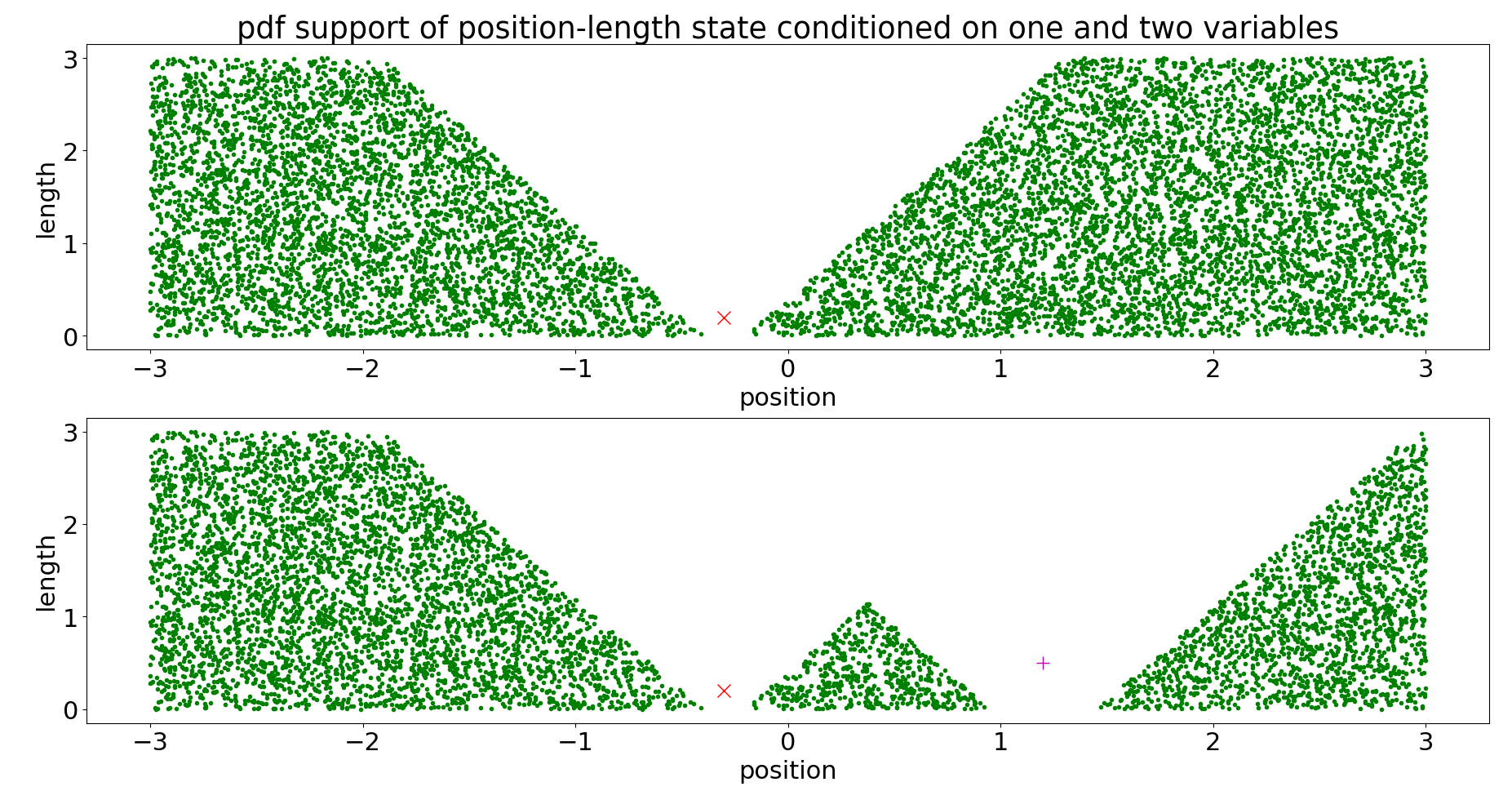} %
 \caption{The support of the pdf of a target's position-length state when conditioned on one other target (top) and on two other targets (bottom). Axes are limited between -3 and 3.}
\label{fig:conditional}
\end{figure}

For point targets, the support is the plane excluding the diagonal line which has a measure zero. Thus, Bayesian inference, which involves Lebesgue integrals, is not affected. We can therefore simply ignore the exclusion and treat the whole space as the support of the relevant pdf. Not so for extended targets: The band of exclusion cannot be ignored. As another example, consider the support of the pdf of the position-length state of a target, conditioned on (i) one other target, and (ii) two other targets. These are shown in Figure~\ref{fig:conditional}, where the conditioning targets are shown by (i) a cross and (ii) a cross and a plus. Again, for point targets, the excluded set consists of only finite points, which has a measure zero. But for extended targets, inference using such conditional pdf's is by no means straightforward.

In the absence of a general mathematical tool to deal with such situations, a practical solution might be to constantly check for spatial compatibility. This may make the Collision Likelihood Structure applicable in many applications.

\section{Conclusions} \label{sec:con}
 In multi-target tracking, a data association hypothesis assigns measurements to tracks, and the hypothesis likelihood (of the joint target-measurement associations) is used to compare among all hypotheses for truncation under a finite compute budget. It is often assumed however that an individual target-measurement association likelihood is independent of others, i.e., it remains the same in whichever hypothesis it belongs to. In the case of Track Oriented Multiple Hypothesis Tracking, this leads to a parsimonious representation of the hypothesis space, with a maximum likelihood solution obtained through solving an Integer Linear Programming problem. In Labeled Random Finite Set filters, this leads to an efficient way of obtaining the top ranked hypotheses through solving a ranked assignment problem using Murty's algorithm. In this paper we have presented a Propose and Verify approach for certain Dependent Likelihood Structures, such that the true hypothesis likelihood is evaluated jointly for the constituent track-measurement associations to account for dependence among them, but at the same time that ranking is still obtained efficiently. This is achieved by proposing a candidate ranking under an assumption of independence, and then evaluating the true likelihood one by one, which is guaranteed, for Collision Likelihood Structure and Occlusion Likelihood Structure among others, to not increase from its candidate value, until the desired number of top ranked hypotheses are obtained. We have illustrated the difference this new approach can make using a simple one dimensional example. We also call attention to the lack of fundamental mathematical tools in dealing with spatial compatibility encountered in target tracking problems.

\section*{Acknowledgment}
The author would like to thank Dr. Giancarlo Baldan for insightful discussions.

\bibliographystyle{IEEEtran}
\bibliography{MyBSTcontrol,rfs}

\appendix

\subsection{Simulation and tracking set up} \label{appd:setup}
The fundamental units are meters and seconds, and will be omitted in the following. The nominal positions of two targets are explicitly specified as time functions which can be seen from Figures~\ref{fig:simu1} and \ref{fig:simu2}, such that one follows the other without overtaking. A birth model is placed at the origin with a large covariance and a birth probability of 0.5. This knowledge is assumed by the tracker, but for simplicity of the illustration, the above two targets are deterministically placed into the simulation while other births are disabled. 

The position sensor has a measurement error standard deviation of 1.0, and a Probability of Detection of 0.99. It has a bounded uniform Poisson clutter model between -50 and 150, with an intensity of $5 \times 10^{-3}$. The coarse gating parameter\cite{ChenL21_merge} is set at 20.0.

The tracker assumes that targets follow a Near Constant Velocity model, with the continuous process noise parameter $q$ being 1.0. The probability of target survival is nearly 1.0. The $\delta$-GLMB filter takes the factored form as presented in \cite{ChenL21_merge}, and the maximum number of hypos per factor is conservatively set at 100, although a smaller value can achieve similar performance.

The tree diagram shown in Figure~\ref{fig:tree} traces the current set of hypos back 5 generations to include in the plot, at each iteration in the debug mode.

\subsection{Notations in the hypo tree} \label{appd:notation}
The following notations are used for a tree diagram such as that in Figure~\ref{fig:tree}, although not everything is present in that figure. For details, see \cite{ChenL21_merge}.
\begin{itemize}
 \item Each ellipse or diamond represents a hypothesis, with the latter being for the current frame of update. The rendering of the tree with the nominal root node called ``head'' is such that hypotheses with a shorter history are located more to the left, as is the case in the lower left corner. 
 \item Each hypothesis shows a listing of comma delimited pairs such as {\tt {\textcolor{red}{12}}.34}, where the red number before the decimal point is the Track ID, while the black number is the Measurement ID and may be absent for missed detections. The weight of the hypothesis is shown on the incoming arrow.
 \item Each hypothesis also shows on the second line a gray number that is the Frame Number of the measurement set for its update. For the diamond, this number is followed by a colon and another number, which is the sequence number of the factor that contains this hypothesis. 
 \item The birth of a hypothesis is also shown with an arrow from the ``head'' node, and an orange integer on it denoting the frame number. 
 \item A simple algorithm is used to obtain an estimate of the tracks at each time: For each factor, the hypothesis with the largest weight defines the tracks. Such hypotheses are colored with a blue border.
 \item When merging happens, the weights are shown in the color red.
 \item When splitting happens, those hypotheses that gate with the measurements have their weights shown in green, while those that don't, in blue. Because of the many-to-many relationships between the ``before'' and ``after'' nodes, the chosen parent-child relationship is only for simple viewing as a tree. In other words, we have to aggregate all the green and blue weights to understand the underlying operation. 
\end{itemize}
\begin{itemize}
 \item Each ellipse or diamond represents a hypothesis, with the latter being for the current frame of update. The rendering of the tree with the nominal root node called ``head'' is such that hypotheses with a shorter history are located more to the left, as is the case in the lower left corner. 
 \item Each hypothesis shows a listing of comma delimited pairs such as {\tt {\textcolor{red}{12}}.34}, where the red number before the decimal point is the Track ID, while the black number is the Measurement ID and may be absent for missed detections. The weight of the hypothesis is shown on the incoming arrow.
 \item Each hypothesis also shows on the second line a gray number that is the Frame Number of the measurement set for its update. For the diamond, this number is followed by a colon and another number, which is the sequence number of the factor that contains this hypothesis. 
 \item The birth of a hypothesis is also shown with an arrow from the ``head'' node, and an orange integer on it denoting the frame number. 
 \item A simple algorithm is used to obtain an estimate of the tracks at each time: For each factor, the hypothesis with the largest weight defines the tracks. Such hypotheses are colored with a blue border.
 \item When merging happens, the weights are shown in the color red.
 \item When splitting happens, those hypotheses that gate with the measurements have their weights shown in green, while those that don't, in blue. Because of the many-to-many relationships between the ``before'' and ``after'' nodes, the chosen parent-child relationship is only for simple viewing as a tree. In other words, we have to aggregate all the green and blue weights to understand the underlying operation. 
\end{itemize}

\end{document}